# Structural and magnetic evidence of confined strain fields in GaMnAs grown on ordered arrays of zero-dimensional nanostructures

X. Martí[1], T. Cechal[1], L. Horák[1], V. Novák[2], K. Hruška[2], Z. Výborný[2], T. Jungwirth[2,3], and V. Holý[1]

[1] Department of Condensed Matter Physics, Faculty of Mathematics and Physics, Charles University, Ke Karlovu 5, 121 16 Prague 2, Czech Republic
[2] Institute of Physics ASCR, v.v.i., Cukrovarnická 10, 162 53 Praha 6, Czech Republic
[3] School of Physics and Astronomy, University of Nottingham, Nottingham NG7 2RD, United Kingdom

We prepared $Ga_{0.95}Mn_{0.05}As$ films on top of periodic arrays of InAs quantum dots. X-ray diffraction reveals periodically strained films, commensurate to substrate's patterning. The dots produce a tensile strain in GaMnAs while between the dots strain is compressive. Our experiments confirm that the average tensile strain in the film increases with decreasing dots separation. This trend in strain is accompanied by an increase of the out-of-plane magnetization component familiar from the established relation between strain and magnetic anisotropy in GaMnAs films. Our work provides a new route for controlling magneto-crystalline anisotropies in GaMnAs on a nanometer scale.

Materials where the magnetic properties are controlled using electric fields constitute a very active field of research [1]. Mn-doped gallium arsenide ($Ga_{1-x}Mn_xAs$) is a semiconductor where ferromagnetic interactions are mediated by holes [2] which directly implies a connection between the magnetic and electrical properties. Its exploitation has led to observations of the electrically induced rotation of magnetization [3], the control of $T_C$ [4] or, very recently, the reversible *on* and *off* switching of the magnetization [5]. These manipulations of the magnetization are often done via the magnetic anisotropy which, for instance, can be tuned by epitaxial strain [6]. While the link between magnetic anisotropy and strain has been thoroughly investigated either by inserting buffer layers [6, 7] or patterning GaMnAs layers to 1-dimensional stripes [8-10], GaMnAs films with zero-dimensional modulation of the epitaxial strain have been only attempted by growing GaMnAs in Stranski-Krastanov mode [11] or on top of InAs quantum dots (QDs) [12].

In Fig. 1 we plot numerical finite-element simulations which show that the in-plane epitaxial strain in the GaMnAs layer is alternatively tensile and compressive on top and between the QDs, respectively, thus showing a path how to confine regions of selected magnetic anisotropy. However, in the former studies the correlation between strain and magnetic properties has not been established, presumably due to the scatter of experimental dot sizes and separations which prevented an accurate determination of the strain fields in the GaMnAs film.

In this letter we report a study of GaMnAs epilayers on top of periodic arrays of InAs QDs grown on pre-patterned GaAs substrates. Satellites around GaMnAs reflections have been observed in X-ray diffraction experiments using synchrotron light. By comparing the intensity of the satellites, we conclude that the magnetic layer is strained periodically and commensurate to the underlying dots array. Particular attention is devoted to discriminate the unavoidable chemical contrast caused by the surface roughness. We show that the average tensile strain can be tuned by changing the QD spacing. Finally, we address the correlation of strain and magnetic properties. We conclude that the out-of-plane magnetization component is increased by increasing the average in-plane tensile strain.

The materials were grown by molecular beam epitaxy (MBE) on GaAs(001) substrates. Prior to growth, four separate fields of 2x2 mm$^2$ areas were defined on the GaAs(001) susbtrate by e-beam lithography and dry etching. Each field contains a rectangular grid of pits of approximately 40 nm diameter and 10 nm depth. In the individual fields, the pit spacing was set to 100, 120, 140, and 160 nm. First, a GaAs layer equivalent to 40 monolayers (according to previous callibration of growth rate) was grown at 590ºC aiming to reduce random surface roughness. Second, the equivalent of 2.5 monolayers of InAs were deposited at 500°C. These growth conditions favour the formation of arrays of InAs quantum dots [15] matched to the pre-patterned pit grid [16], as confirmed by our atomic force microscopy images shown in the inset 2 of Fig. 1. Finally, the equivalent of a 60 nm thick $Ga_{0.95}Mn_{0.05}As$ was deposited at 250°C. Note that the low temperature MBE growth of GaMnAs is used not only to prevent the formation of unintentional precipitates in the magnetic film but also to reduce the mobility of the incoming adatoms which enhances the imprint of the bottom InAs QD morphology in the GaMnAs film (see inset 3 in Fig. 1). After the growth, the substrate was cut into 4 separate pieces containing always only the array with selected pit spacing. In the following, different samples will be identified with the different QD spacings in the array.

We now turn to the X-ray diffraction experiments performed at ESRF-ID10B. Figs. 2a and 2b show reciprocal space maps for the 140 nm sample around GaAs(004) and GaAs(224) reflections, respectively. Note that the data are shifted to place the GaAs peak at the origin of coordinates. Of relevance here is the presence of satellites around the GaAs and GaMnAs peaks in both panels. For the sake of clarity, two horizontal profiles along the indicated horizontal arrows have been plotted in panels (c) and (d) in Fig. 2. Satellites (highlighted by vertical arrows) are located at $\pm\Delta Q_x$, symmetrically around the position of GaAs, and correspond to an in-plane periodicity of $2\pi/\Delta Q_x =$ 141±4 nm, in agreement with the lithography patterning of the substrate. For the remaining samples we determined the in-plane periodicity of 100±2 nm, 121±5nm, and 159±3 m, respectively, in excellent agreement with the nominal parameters of the samples. Thickness oscillations around the GaMnAs peak shown in Fig. 3a indicate a thickness of 65±2 nm, again in good agreement with the nominal thickness of the magnetic film.

We now focus on the intensity of the observed satellites. In an array with lateral period *L,* the amplitude of the *n*-th lateral satellite around the reciprocal lattice point ***h*** is given by [14]:

$$E_{nh}(q_z) \sim \frac{1}{L} \int_{-\frac{L}{2}}^{\frac{L}{2}} dx \int_{-\infty}^{\infty} dz \, \chi_h(x,z) e^{-i\mathbf{h} \cdot \mathbf{u}(x,z)} e^{-i(nG+q_z z)}, \, G = 2\pi/L \quad (1)$$

where $\mathbf{u}(x, z)$ is the elastic displacement with respect to the fully relaxed atomic position and $\chi_h(x, z)$ are the $\mathbf{h}$-Fourier coefficients of the crystal polarizability. While the vertical component $u_z(x,z)$ is symmetric, i.e. $u_z(x,z) = u_z(-x,z)$, due to the mirror symmetry of the dot shape, the in-plane component $u_x(x,z)$ of $\mathbf{u}(x,z)$ is antisymmetric, i.e. $u_x(x,z) = -u_x(-x,z)$. (For each atom displaced to the right on one side, there is another equally displaced to the left on the other side when measured form the center of the QD.) It can be shown [14] that the diffracted intensity ($I \sim |E_{1h}(q_z)|^2$) of the first satellites will be identical, $I(q_x) = I(-q_x)$, only if $\mathbf{h} \cdot \mathbf{u} = 0$ or if $\mathbf{h} \cdot \mathbf{u}(x,z) = \mathbf{h} \cdot \mathbf{u}(-x,z)$. This occurs if either $\mathbf{h}$ is vertical (like in the (004) reflection) or if $\mathbf{u} = 0$. It turns out that, if the elastic strain is present, the intensity of the satellites located at $\pm\Delta Q_x$ should differ in the asymmetrical (224) reflection but should be identical in the symmetrical (004) case. Data in Figs. 3c and 3d clearly support this expectation. Identical results are obtained for all measured samples and we can conclude that there is a periodic modulation of the epitaxial strain in the GaMnAs epilayers.

We now address the question whether the ratio of tensile and compressively strained parts of the material changes by reducing the QD spacing. We take the location of the maxima in the $Q_z$ profile in the (004) maps as a signature of the average strain (see Fig. 3b). As the QD spacing is reduced, peaks gradually shift to the right which indicates a progressive contraction of the out-of-plane lattice parameter and, in turn, an expected increase of the in-plane tensile component of the epitaxial strain.

We now turn to the magnetic properties of the films that were inferred from SQUID measurements. In typical GaMnAs epilayers, in-plane tensile (compressive) strain favors the out-of-plane (in-plane) easy magnetic moment orientation [7, 13]. Therefore, at a given temperature and Mn doping, a smaller relative size of the compressively strained part of the GaMnAs epilayer in samples with smaller QD spacing should lead to a larger net out-of-plane component of the magnetization at a given magnetic field. In Fig. 4 we show the field cooled temperature dependence of the out-of-plane magnetization measured at 3 kOe. Given the intricate temperature dependence of the magneto-crystalline anisotropies in GaMnAs [7, 13], we chose 50 K ($m_{50K}$) as the most appropriate temperature for observing the changes in the in-plane vs. out-of-plane anisotropy. In the inset of Fig. 4 we plot $m_{50K}$ as a function of the $Q_z$ maximum of GaMnAs(004) reflection which accounts for the average tensile strain. In agreement with our expectations, the data show a clear correlation indicating that the out-of-plane magnetization increases with increasing in-plane tensile strain. We have therefore established a clear link between local strains controlled on the nano-scale and changes of the averaged magnetic anisotropies in GaMnAs ferromagnetic semiconductor films. This provides a strong motivation for future studies using local magnetic probes which will resolve the length-scale of magnetic anisotropy variations and will determine the limits of technique for controlling magnetic properties in GaMnAs nano-magnets.

In summary, we have grown GaMnAs on an array of ordered InAs QDs. Using x-ray diffraction we have concluded that the epitaxial strain in GaMnAs is periodically modulated and its period can be controlled by pre-patterning the substrate. We have shown that by reducing the QD spacing the average amount of material under in-plane tensile strain increases which is correlated with an increasing out-of-plane magnetization. Our results provide guidelines for the research towards control of the magnetic anisotropy in zero-dimensional nanostructures for high-density spintronic devices.

The work has been supported by the EC projects NAMASTE (No.214499) and SemiSpinNet (No. 215368) and by the Academy of Sciences and Ministry of Education of the Czech Republic (No. AV0Z10100521, No. KAN400100652, No. LC510, Praemium Academiae). We acknowledge the staff at ID10B beamline (ESRF) and B. Bittová (Charles University in Prague) for AFM measurements.

**Figure captions**

Fig. 1: Sketch of the GaMnAs/InAs/GaAs nanostructure. The GaAs substrate (1) is etched before the subsequent growth of InAs quantum dots (2). Finally, GaMnAs is grown on top (3). Surface topography in each step is shown at bottom. GaMnAs in-plane isostrain lines are plotted showing the alternation of compressive (side) and tensile (center) strain.

Fig. 2: Reciprocal space maps around the (a) GaAs (004) and (b) GaAs(224) reflections for the sample with 140 nm dot spacing. Profiles along the horizontal lines in panels (a) and (b) are plotted in panels (c) and (d), respectively. Satellites are signaled by vertical arrows, being symmetric in the (004) reflection and asymmetric in the (224) reflection.

Fig. 3: (a) symmetrical $Q_z$ scan for the 140 nm sample displaying the (004) reflection by GaAs substrate and GaMnAs film. (b) Zoom around the GaMnAs(004) reflection for the complete set of samples.

Fig. 4: Temperature dependence of the out-of-plane magnetization measured after FC at 30 K at 3 kOe. Inset shows the magnetization at 50 K as a function of the $Q_z$ maxima in Fig. 3 to illustrate the correlation between magnetic properties and epitaxial strain.

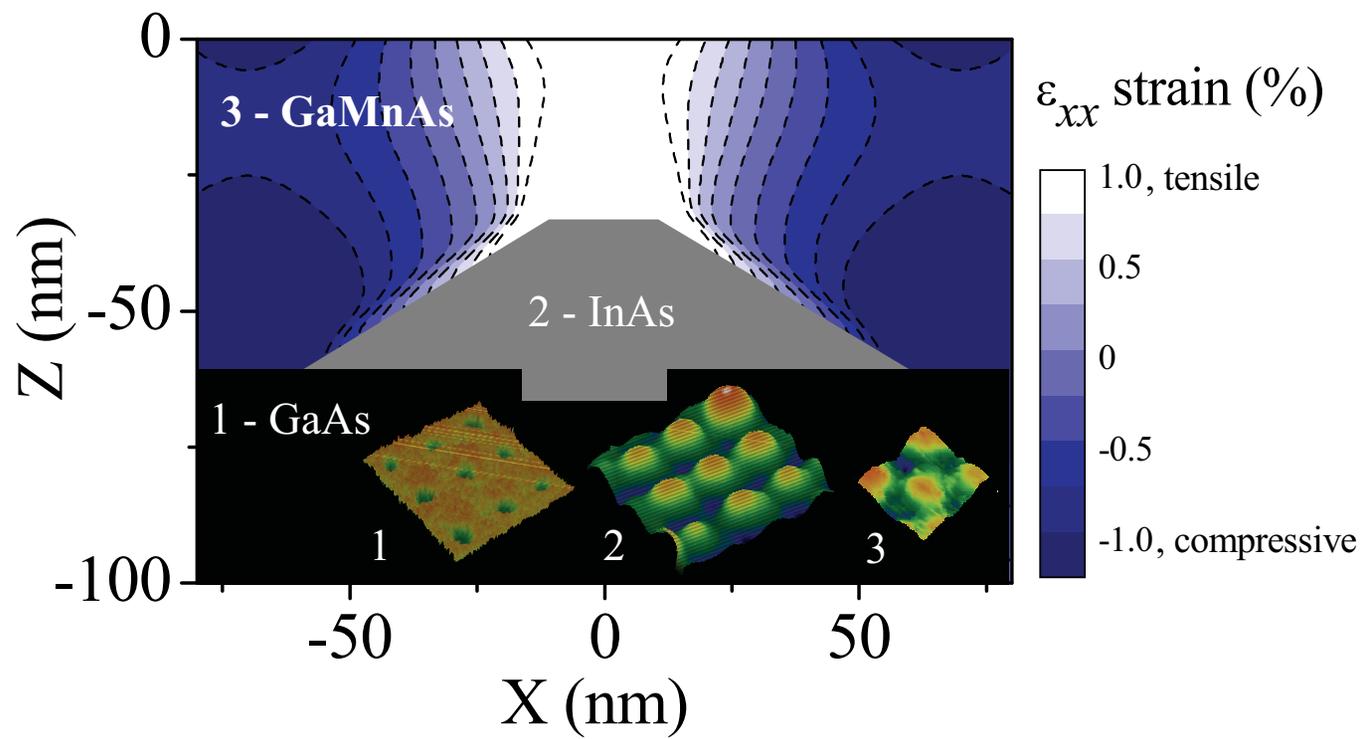

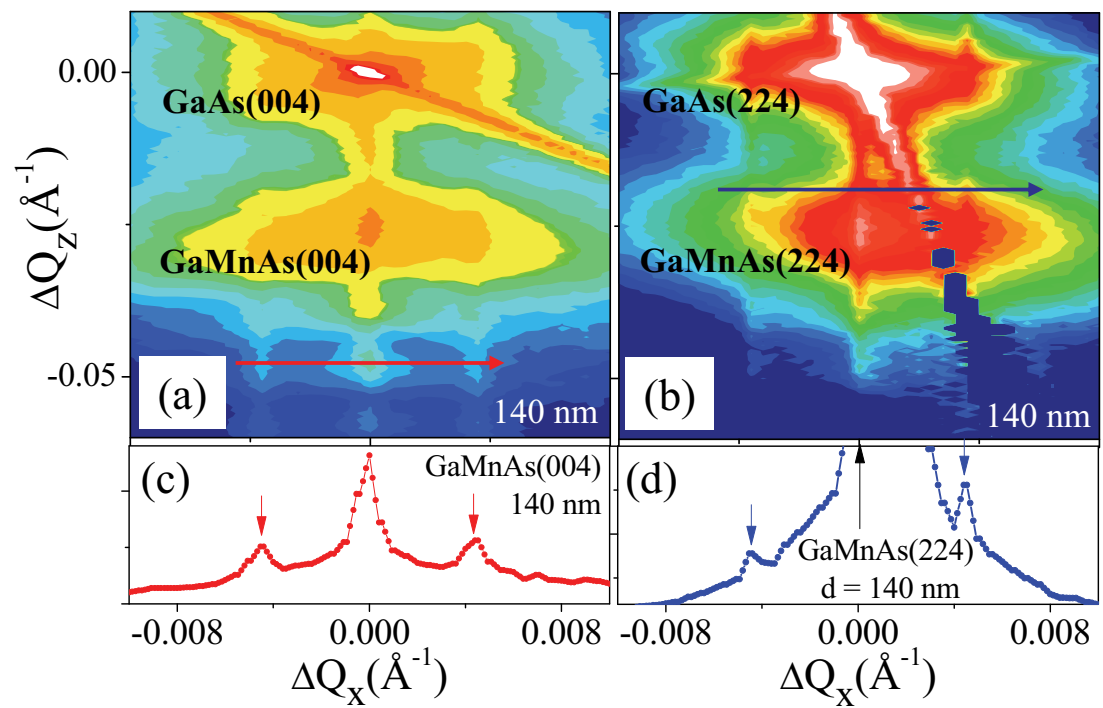

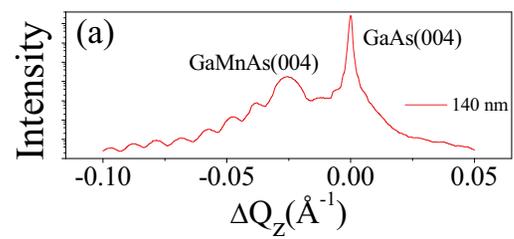
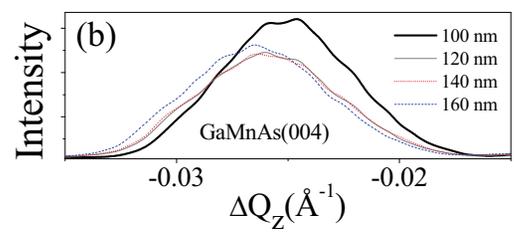

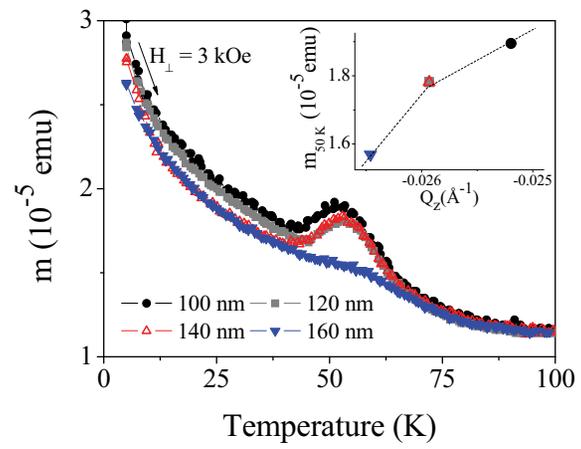